# Large-scale Graphitic Thin Films Synthesized on Ni and Transferred to Insulators: Structural and Electronic Properties

Helin Cao[1,2,#], Qingkai Yu[3,#], Robert Colby[2,4], Deepak Pandey[1,2], C. S. Park[5], Jie Lian[6], Dmitry Zemlyanov[2], Isaac Childres[1,2], Vladimir Drachev[2,7], Eric A. Stach[2,4], Muhammad Hussain[5], Hao Li[8], Steven S. Pei[3] and Yong P. Chen[1,2,7,*]

[1]Department of Physics, Purdue University, West Lafayette, IN 47907 USA

[2]Birck Nanotechnology Center, Purdue University, West Lafayette, IN 47907 USA

[3]Center of Advanced Materials and Department of Electrical and Computer Engineering, University of Houston, Houston, TX 77204 USA

[4]School of Materials Engineering, Purdue University, West Lafayette, IN 47907 USA

[5]SEMATECH, 2706 Montopolis Dr., Austin TX 78741 USA

[6]Department of Mechanical, Aerospace and Nuclear Engineering, Rensselaer Polytechnic Institute, Troy, NY 12180 USA

[7]School of Electrical and Computer Engineering, Purdue University, West Lafayette, IN 47907 USA

[8]Department of Mechanical and Aerospace Engineering, University of Missouri, Columbia, MO 65211 USA

# Equally contributing authors.   * Corresponding author. E-mail: yongchen@purdue.edu

**Abstract**: We present a comprehensive study of the structural and electronic properties of ultrathin films containing graphene layers synthesized by chemical vapor deposition (CVD) based surface segregation on polycrystalline Ni foils then transferred onto insulating $SiO_2$/Si substrates. Films of size up to several mm's have been synthesized. Structural characterizations by atomic force microscopy (AFM), scanning tunneling microscopy (STM), *cross-sectional* transmission electron microscopy (XTEM) and Raman spectroscopy confirm that such large scale graphitic thin films (GTF) contain both thick graphite regions and thin regions of few layer graphene. The films also contain many wrinkles, with sharply-bent tips and dislocations revealed by XTEM, yielding insights on the growth and buckling processes of the GTF. Measurements on mm-scale back-gated transistor devices fabricated from the transferred GTF show ambipolar field effect with resistance modulation ~50% and carrier mobilities reaching ~2000 $cm^2$/Vs. We also demonstrate quantum transport of carriers with phase coherence length over 0.2 μm from the observation of 2D weak localization in low temperature magneto-transport measurements. Our results show that despite the non-uniformity and surface roughness, such large-scale, flexible thin films can have electronic properties promising for device applications.

Graphene has attracted tremendous interest for its potential to enable new carbon-based nanoelectronics and other nanotechnology applications[1]. Numerous methods have been developed for the production of graphene and graphene-based materials[1-4]. While the study of *electrically isolated* graphene was quite recent[1,2], graphene has been synthesized by chemical vapor deposition (CVD) and surface segregation on various transition metals and other conducting substrates for several decades[5]. One of the most commonly used substrates has been Ni and graphene layers (down to monolayer) grown on Ni have been extensively studied[5,6]. On the other hand, it was only very recently that such graphene layers were transferred (by removing the Ni substrates) to insulating substrates[7-10], enabling electronic applications. This renewed interest in graphene growth on Ni has also been driven by the promise for large-scale production of graphene[4,7-10]. While high quality monolayer graphene can be grown *locally* on Ni[7-10], over large areas, however, the films grown on Ni demonstrated so far[7-12] generally have non-uniform thickness and surface roughness (eg. wrinkles) as a result of the growth process[7-12]. A good understanding of the electronic properties of such large scale graphitic thin films (GTF) in the presence of structural non-uniformity and roughness is important for their applications as flexible and transparent electronic materials. This may also be relevant for graphene-based materials synthesized by many other methods[2,3] (e.g. numerous solution-based chemical approaches[13-16]) which often give large but non-uniform graphitic thin films made of connected regions containing different number of graphene layers. In addition, many structural features or defects, such as wrinkles, need to be better understood on a microscopic level in order to clarify how they are formed, and how they may affect the mechanical as well as electronic properties of the large scale films[9,11,12].

In a previous publication (Ref. 7), we reported synthesis of high quality, transferrable graphene layers using a CVD-based segregation approach on polycrystalline Ni foils. In this paper, we present a comprehensive study of the structural and electronic properties of thus-synthesized *large-scale* GTF *after* transfer onto insulating substrates ($SiO_2$/Si). We performed various structural characterizations to examine the crystalline quality of the transferred films and confirmed the thickness non-uniformity and presence of surface roughness (eg. wrinkles). A *cross-sectional* TEM (XTEM) analysis was used to examine how the transferred graphene layers relate to the underlying substrate, and how wrinkles may have formed by buckling of the thin film. We performed various electronic and magneto transport measurements in *large* (mm-scale) samples. We observed ambipolar field effect and carrier mobilities over ~2000 cm$^2$/Vs. We also observed weak localization and extracted information on the scattering and quantum coherent transport of carriers. Our results show that despite the structural irregularities, our large-scale graphitic films can have excellent electronic properties for potential devices applications.

Our method to grow graphene layers on low-cost, commercially available polycrystalline nickel (Ni) foils has been described in detail previously[7] (see also Appendix). The growth is predominantly a non-equilibrium surface segregation process by controlled cooling, conducted in an *ambient-pressure* CVD furnace, and is qualitatively similar to other recent studies using evaporated Ni films as growth substrates[8-10]. Our previous studies[7] have shown that one to multiple layers of graphene of high quality can be grown by such a method and transferred to other substrates. The number of graphene layers grown is found to depend sensitively on many experimental parameters, including the pressure and composition of the precursor gas, growth time, substrate temperature, cooling rate and so on[5,7-12]. As a result, the thickness of the film is generally non-uniform at large scale[7-12]. In our case, we can often find regions as thin as a few (<6) layers of graphene[7] and as thick as tens of layers in the same sample. Transferring as-grown GTF from metals to insulator substrates is a critical step for fabricating electronic devices. For the current work, we have used a simple method of substrate transfer (differently from what we used in Ref. 7), without the need of any auxiliary adhesive or supportive coatings[7-10]. The Ni substrate with as-grown films is placed in an acid solution, where the GTF detaches from the Ni and floats on top of the liquid surface. This is demonstrated in Fig. 1a for a large film, close to 1 cm in size and largely transparent. The film can then be simply skimmed out by the insulator substrate onto which it is transferred (Fig. 1b). The type of insulator substrate used in this work is 300 nm SiO$_2$ thermally-grown on a doped Si wafer. Fig. 1c shows an optical microscope image of a synthesized GTF transferred to such a SiO$_2$/Si substrate. We can clearly see both the color variation (related to the thickness non-uniformity[17-18]) and the wrinkles/ridges on the film.

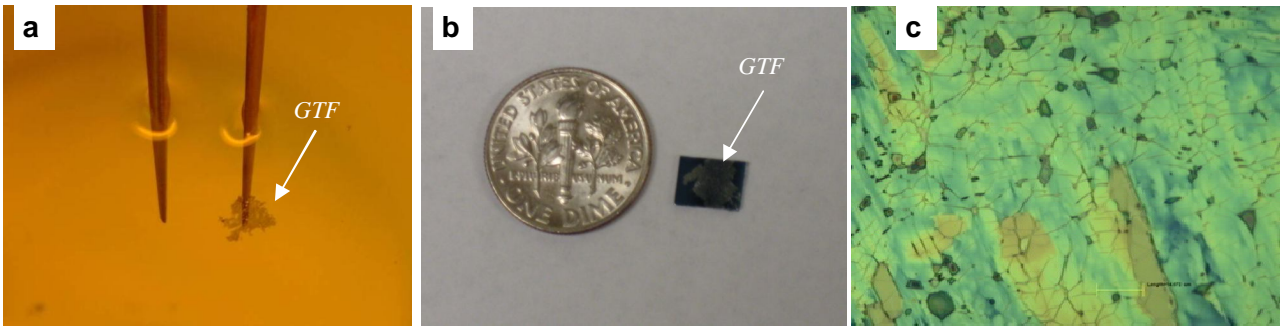

Fig. 1. Transfer of large scale graphitic thin film (GTF) to insulator substrate. (a) Photograph of a large-size film floating on the surface of the acid solution (HNO$_3$) used to etch the film off its growth substrate (nickel). A pair of tweezers is placed behind the largely-transparent film. (b) Photograph of the film in (a) transferred onto a SiO$_2$/Si wafer. (c) Optical microscope image of a transferred film. The color variation, related to thickness non-uniformity, and the wrinkles on the film are clearly visible (the scale bar is 15 µm). A relatively thick film is shown in (a,b,c) for better visibility.

Results of structural characterizations using several different experimental tools of our synthesized GTF after transferring onto SiO$_2$/Si are summarized in Fig.2. Fig. 2a shows an atomic force microscopy (AFM) image near the edge of a transferred GTF and Fig. 2b shows the cross sectional height profile corresponding to the line in Fig 2a. The step from the SiO$_2$ surface to the GTF here has a height <~ 1 nm, indicating only 1-2 layers of graphene[8, 19]. The AFM image also shows height fluctuations due to wrinkles, which will be discussed later. Fig. 2c shows representative Raman spectra measured (with a 532nm excitation laser) from several different locations in a typical transferred GTF. The spectra 1-4 (from top to bottom, offset for clarity) are displayed in the order of decreasing intensity ratio of the "2D" band (~2700 cm$^{-1}$) to "G" band (~1580 cm$^{-1}$), corresponding to increasing thickness (number of layers) at the locations measured. In particular, the spectrum #1 has a 2D/G intensity ratio ~2, likely corresponding to only 1-2 layers of graphene[7-12, 20-22]. Compared to graphene exfoliated on $SiO_2$[20-22], it can be more difficult to determine the exact layer thickness of our samples from the Raman spectrum due to possible stacking disorder in the first few layers of CVD grown graphene[8-11]. The intensity of the disorder-induced "D" band (~1350 cm$^{-1}$) in our transferred GTF samples is typically no more than ~10% of the "G" band intensity, indicative of good crystalline quality[7-10,20-22]. Fig. 2d and 2e show the cross-sectional transmission electron microscopy (XTEM) images of thick and thin regions in transferred GTF. Such XTEM analysis can provide the most unambiguous determination of any layer thickness by directly counting the graphene layers. It is, however, partially disruptive to the samples and must be performed with care (see Appendix). The thick region in (d) contains many graphene layers, which show good structural quality (even after the partially disruptive TEM processes) and are parallel to the $SiO_2$ substrate underneath the transferred GTF. The thin region shown in (e) contains less than 6 graphene layers. The substantial amount of defects seen are believed to result from the ion beam induced damage inflicted on these relatively vulnerable few graphene layers, and do not necessarily reflect the quality of the GTF before being subject to the partially destructive processes in preparing and imaging TEM cross sections (see Appendix). Fig. 2f shows a high-resolution scanning tunneling microscopy (STM) image of a transferred GTF. The hexagonal pattern of the underlying graphitic lattice is clearly revealed, down to atomic resolution. Such images can contain valuable information about atomic-scale roughness and defects (for example adsorbents and vacancies[23]) and lattice distortions in the graphene layers.

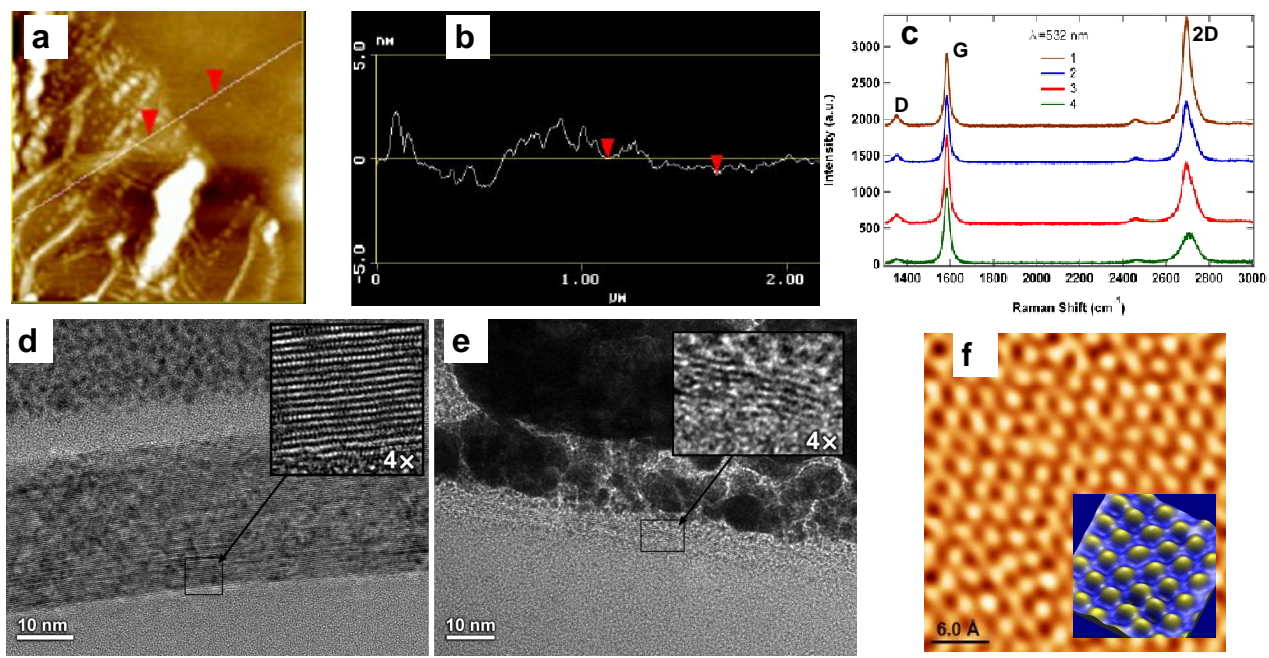

Fig. 2. Characterizations of synthesized GTF transferred on $SiO_2/Si$. (a) Atomic force microscopy (AFM) image near the edge of a transferred GTF. (b) Cross sectional height profile corresponding to the line in (a). The step from the $SiO_2$ surface to GTF has a height <~ 1nm, indicating only 1-2 layers of graphene in this region. (c) Raman spectra (offset for clarity) measured from 4 different locations (with different thickness) on a typical transferred GTF. The spectra were measured with a 532nm excitation laser and a 100X objective lens. (d,e) Cross-sectional transmission electron microscopy (XTEM) images of thick (d) and thin (e) regions. The thin layer between the graphene layers and the Pt protective layer (see Appendix) is likely residual photoresist from sample fabrication. The insets show magnified views near the regions indicated by the arrows. The thinner region in (e) was imaged at a lower voltage (80kV, thus the slightly reduced image quality) than that (300 kV) used in (d) to help prevent electron beam induced damage. A substantial amount of defects seen in these few layers could result from the ion beam induced damage from the partially destructive processes in preparing and imaging TEM cross sections (see Appendix). (f) High resolution scanning tunneling microscope (STM) image in a relatively flat region, revealing the underlying hexagonal graphitic lattice structure. The image mainly resolves atoms from only one sub-lattice due to the stacking order of graphene layers in graphite (multiplayer graphene)[23]. A wavelet-based filter is used to enhance the contrast of atomic images. Inset shows a 1.25 nm x 1.25nm region in magnified 3D view.

Wrinkles are often the dominant topography features on graphitic films segregated on Ni[11,12]. The larger wrinkles in our samples can be visible even in optical microscope images (Fig. 1c). Fig. 3a is a lower resolution STM image of a transferred GTF, showing such wrinkles as tall ridges (height ~60-100 nm) separating μm-scale, relatively flat regions. Although some wrinkles are introduced during the transfer process, many of them are already present before the transfer, as shown in the atomic force microscope (AFM) image (Fig. 3b) of an as-grown film on a Ni substrate. The height of some of the larger wrinkles can reach ~100 nm (inset of Fig. 3b), much larger than those of the thickness fluctuations in the graphene layers or the substrate roughness of the Ni. The wrinkles also appear to be sharply bent, analogous to folded paper. Fig. 3c shows a TEM image of the cross section of an isolated wrinkle on a transferred GTF. The wrinkle is seen to display a *sharp*, approximately-90-degree bend, accompanied by an apparent dislocation line originating from the bend. We have studied a number of such wrinkles in relatively thick regions (containing tens of graphene layers) of both transferred and as-grown films, and have observed qualitatively similar features from all of them. The formation of such wrinkles have been suggested as a mechanism of strain relaxation[11,12] caused by the coefficient of thermal expansion (CTE) mismatch between Ni ($\alpha$ = 12.9~21.0×10$^{-6}$ K$^{-1}$)[24] and graphene (0.7~1.2×10$^{-6}$ K$^{-1}$)[25] during cool-down. For example, the wrinkles could form when graphene layers nucleated from different regions of the Ni coalesce and delaminate from the *contracting* substrate during the cooling-induced segregation process. Defects in the graphene layers may also constitute local weak spots aiding the formation of wrinkles and dislocations. Previous studies[11,12], however, have not revealed the atomically sharp bending and dislocations seen in our high resolution cross-sectional TEM imaging. Such a microstructure analysis of the wrinkles provides valuable insights on the growth and buckling processes of the GTF. We note that wrinkles formed by similar mechanisms of thermal expansion mismatch are also commonly seen on graphitic films grown on other solid substrates, such as those by sublimation of SiC[26]. Studies of thin film buckling and bending have received great attention in recent years, motivated both by practical applications (for example in integrated circuits and flexible electronics)[27,28] of thin film materials and fundamental interest (for example, membrane mechanics)[29]. Smooth and nm-sized ripples (instead of sharply-bent buckles) have been expected and observed in mesoscopic (μm sized) single or bilayer graphene and shown to be essential for the stability of such 2D crystals[30,31]. On the other hand, the large, *sharp* wrinkles we have observed may be characteristic of large-scale and/or thicker films and important for the stretchability and mechanical stability of such films for applications such as flexible electronics[9].

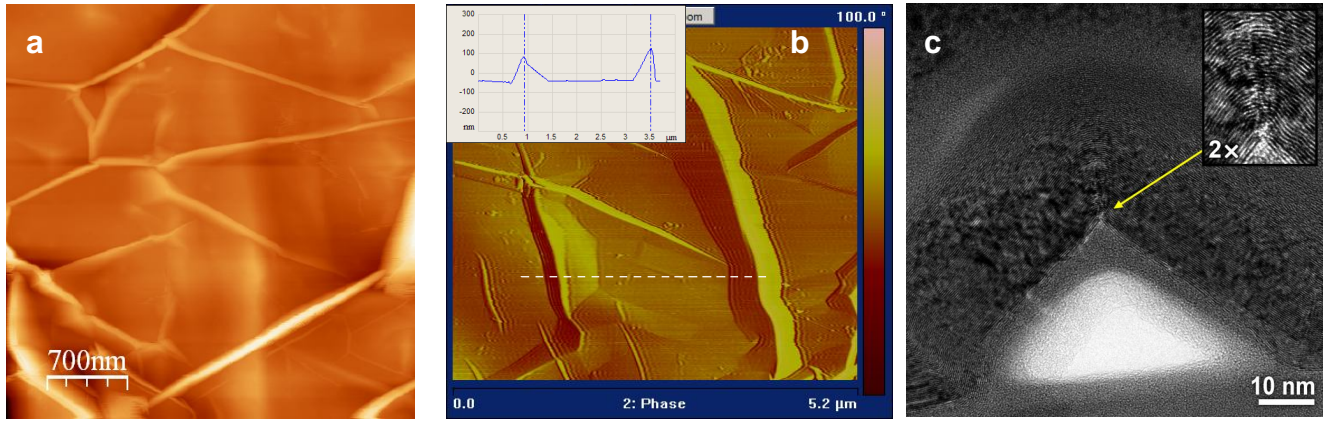

Fig. 3. Characterizations of wrinkles. (a) Low resolution STM image of a transferred GTF, with relatively flat domains separated by tall ridges/wrinkles (tens of nm in height). (b) Atomic force microscope (AFM) image (phase-contrast) of an as-grown film on Ni (before transferring), showing the topography of the wrinkles/ridges on the "crumpled" film. Inset shows cross sectional height profile corresponding to the dashed line in (c), through two large wrinkles. (c) Cross section image of a wrinkle on a transferred film, taken by transmission electron microscopy (TEM). Other disorder features, especially those in the top layers, may be related to beam induced damages from TEM processes and wrinkles along the imaging direction (see Appendix). The film is covered by a thin photoresist and protective amorphous Pt layer. The bright triangular area is due to the void between the wrinkle and the $SiO_2$. Yellow arrow indicates the atomically sharp bending of the wrinkle, with an apparent dislocation line (pointing upward) originated from the bend. The sharp bend is magnified in the inset.

To characterize the electronic properties of the transferred GTF, we have patterned them into relatively large-sized Hall-bars with multiple contact electrodes (Cr/Au). The devices are fabricated by standard optical lithography, plasma etching and metal deposition. An optical microscope image of a typical device is shown in Fig. 4a, and the schematic cross-section in Fig. 4b. We have mainly performed two types of electronic measurements at various temperatures: 1) field effect (transistor) measurements, where the heavily doped Si wafer is used as the back gate; 2) magneto-transport measurements, where a perpendicular magnetic field is applied to the film.

Fig. 4c shows a 4-terminal resistance measured in a device ("A") at low temperature ($T$ = 10 K), where the back gate voltage ($V_{gate}$) is varied between -90 V to 60 V. An ambi-polar field effect is evident, where the resistance can be modulated by more than 50 %, with its peak ("charge-neutral point") occurring at ~10 V. Similar field effects are observable up to room temperature, though at low $T$, a larger range of $V_{gate}$ can be accessed without gate leakage. The insets (Fig. 4d and 4e) show Hall effects measured at $V_{gate}$ = -20 V and 40 V respectively, with opposite signs of Hall slope observed. Such a sign-change in the Hall effect directly shows that our film can be electrically doped from p-type (Fig. 4d) to n-type (Fig. 4e). The extracted carrier mobility for $V_{gate}$ far away from the charge neutral point is ~2000 cm$^2$/Vs. We note that despite the macroscopic size of our sample (with significant thickness non-uniformity and many wrinkles), its mobility is not much lower than those of typical exfoliated[19,32] (uniform in thickness) and CVD-grown[8,9] few-layer graphene flakes of much smaller size and less roughness. We also note that the *ambipolar* field effect we observed are similar to those previously observed in ultrathin 2D graphitic systems (where electronic transport in the field effect is believed to be dominated by only few graphene layers[19,33,34]), and are likely controlled mostly by the thin regions of our samples. Such an ambipolar field effect are important for various rf/high frequency electronic applications[35,36,37]. It has also been pointed out that for RF devices, even moderate mobilities and field effect "on/off" modulation ratio such as those observed in our samples (~50 %) can already be quite valuable[35,36].

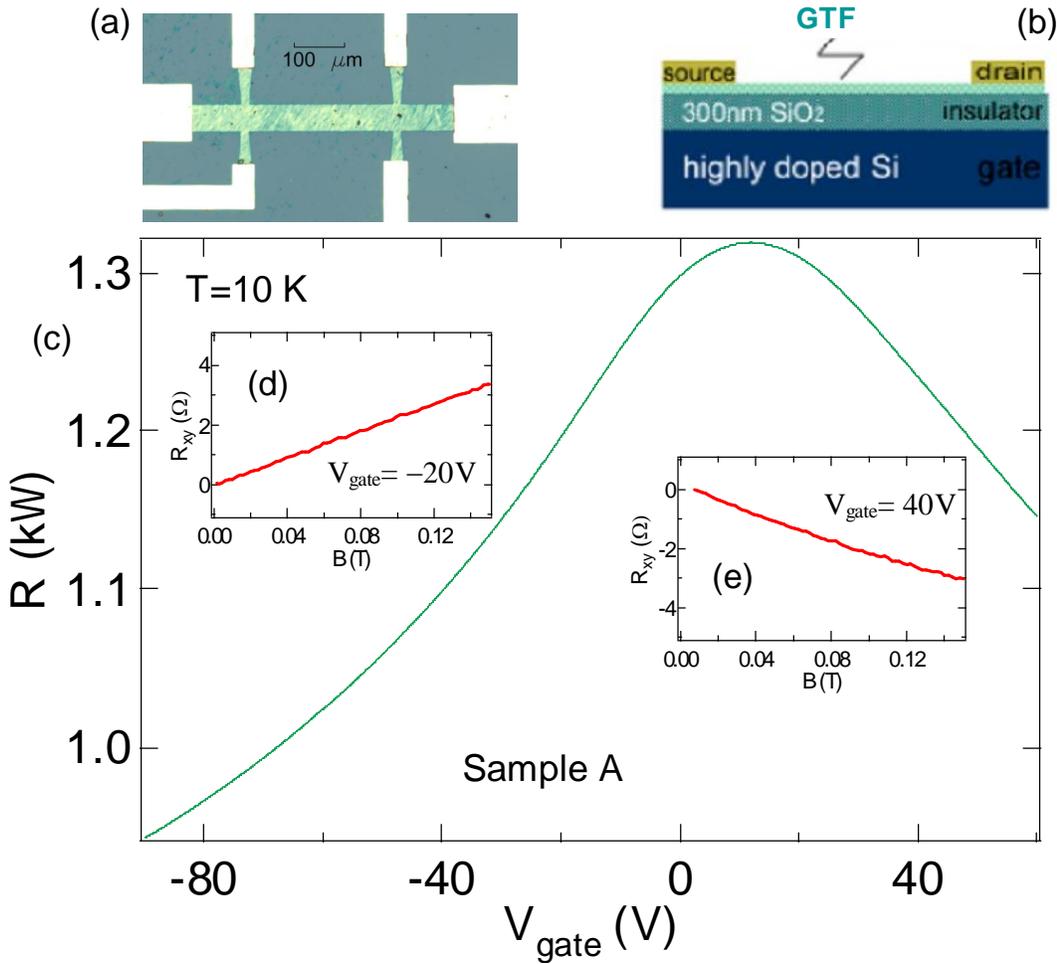

Fig. 4. Electronic devices and ambipolar field effect. (a) Optical microscope image of a typical field effect transistor (FET) device (top view) fabricated from a large scale transferred film, patterned into the Hall bar shape by photolithography. (b) Schematic diagram (not to scale) of the cross section of a FET device. (c) Four-terminal resistance as a function of back gate voltage ($V_{gate}$), showing the electric field effect. (d) Hall effect (Hall resistance as a function of perpendicular magnetic field) measured at $V_{gate} = -20$ V, showing a positive sign and predominantly p-type carriers. (e) Hall effect measured at $V_{gate} = 40$ V, showing a negative sign and predominantly n-type carriers. Data in (c, d, e) are measured in the same sample ("A") at a temperature of 10 K. The extracted mobilities for the majority carriers reach ~2000 $cm^2$/Vs.

Fig. 5a shows the magneto-resistance $\Delta R_{xx}(B) = R_{xx}(B) - R_{xx}(B = 0\,\text{T})$ measured between -0.4 T and 0.4 T in a device ("B") at several temperatures. At low temperature (eg., $T = 0.5$ K), we observe a pronounced low field *negative* magnetoresistance (with a peak in $R_{xx}$ at B = 0 T), which weakens at elevated temperatures and disappears at sufficiently high $T$ (>~ 20 K). Such features are characteristic of so called "weak-localization" (WL)[38], and have been observed in many other graphitic systems[2,32,39-44].

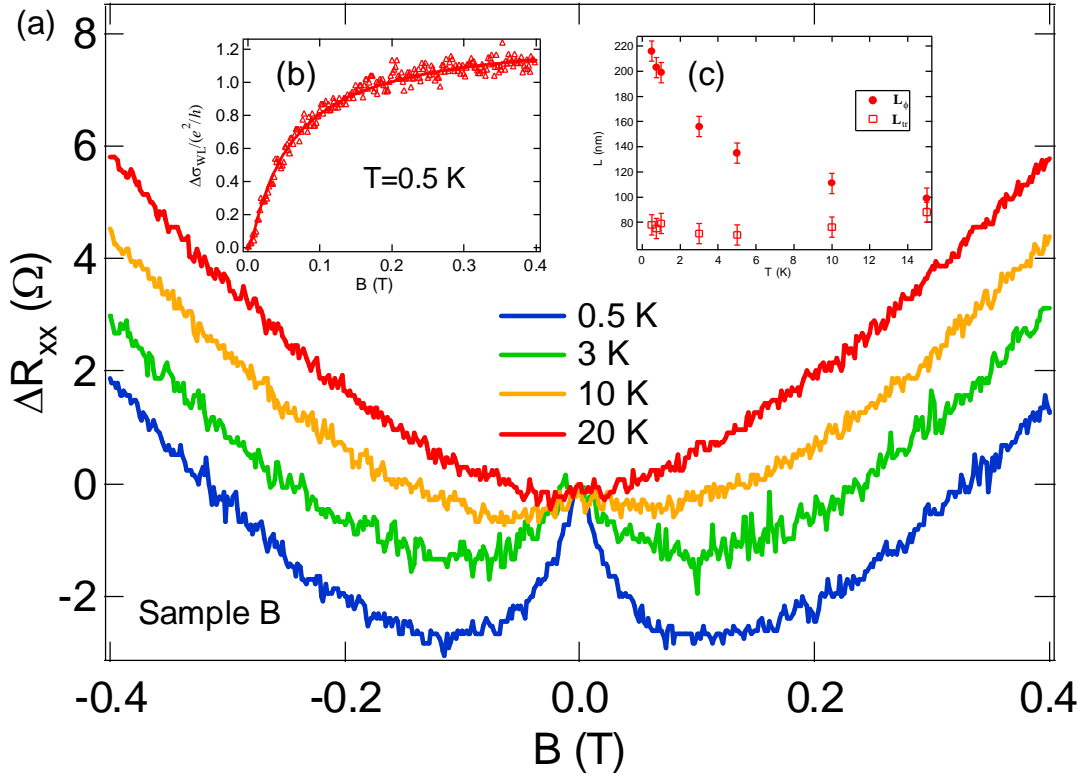

Fig. 5. Quantum transport and weak localization. (a) Magnetoresistance $\Delta R_{xx}(B)= R_{xx}(B)−R_{xx}(0T)$ at several temperatures. $R_{xx}$ is the four-terminal (longitudinal) resistance. (b) Magnetoconductivity (normalized by $e^2/h$) $\Delta\sigma(B)= \sigma(B)−\sigma(0T)$ at 0.5 K. Empty circles are measured data (after subtracting a high temperature background to extract the WL correction). Solid line is the fit using the 2D WL theory[38] (see text). (c) Extracted phase coherence length (filled circles) and transport scattering length (empty squares) as a function of the temperature.

WL arises from the constructive quantum interference between time-reversed multiple-scattering trajectories (within a length scale $L_\varphi$ that the electron wave function is phase coherent) that enhances the probability of electron localization (and thus also the electrical resistance). Such a WL can be destroyed by a magnetic field (which breaks the time-reversal symmetry) or by high temperature, giving rise to the observed negative magnetoresistance at low $T$. We found that our data can be well described by the standard WL theory for a *2D* diffusive metal[38] (see Appendix). Excellent fit (an example is shown in inset 5b) has been obtained for the entire range of measured data for all temperatures at which WL has been observed, using $L_\varphi$ (the afore-mentioned phase coherence length) and $L_{tr}$ (the transport scattering length) as two fitting parameters. Inset 5c shows $L_\varphi$ and $L_{tr}$ thus extracted and plotted as a function of the temperature. $L_{tr}$ is found to be largely temperature-insensitive in the measured range, and its value is on the same order of magnitude as the transport mean-free path (~30-60 nm) extracted from the mobilities. The phase coherence length $L_\varphi$ steadily rises as $T$ is lowered, reaching above 0.2 μm at low $T$ (0.5 K). The $L_\varphi$ is much smaller than the size (hundreds of μm) of our devices, consistent with the diffusive[38] limit of the electronic transport. On the other hand, we envision that phase coherent or ballistic transport may be explored in much smaller submicron devices fabricated from our CVD-grown GTF. This may lead to novel quantum devices[45,46], where phase coherent electron transport is utilized for new device functionalities or improved performance.

WL has received strong attention in recent studies of both exfoliated[32,40-43] and epitaxially grown (on SiC)[2,44] graphene systems. A host of important information valuable to develop and improve graphene-based electronics, including information about the scattering and quantum transport of carriers, as well as the disorder in the samples, can be obtained by careful analysis of magnetotransport behavior such as WL. In our case, the atomic scale defects that we imaged, such as the atomically sharp bending and dislocation associated with the wrinkles, could be part of the disorder that results in WL. On the other hand, the presence of non-uniformity and roughness (many wrinkles) in our large sized samples does *not* appear to be detrimental for the device performance, gauged by, for example the field effect and mobility (comparing to smaller samples with less roughness). The wrinkles have in fact been suggested to be beneficial for the mechanical flexibility[9] of the large-size film.

Much of the active research concerning graphene is inspired by the remarkable properties discovered in *uniform* few-layer (especially monolayer) graphene[1], leading to a focus on the growth of large-scale graphene with *uniform* thickness. This is undoubtedly a noble goal, with rapid advances made recently[47,48]. On the other hand, many of the existing graphene production methods (including numerous low-cost, solution-based chemical approaches[3,13-16]) yield large but *non-uniform* films consisting of an electrically connected mixture of both thick and thin regions, as is the case with the graphitic thin films grown and transferred from Ni studied here. Besides interesting structural and mechanical properties of such large-scale films, we have demonstrated a number of electronic properties previously studied in uniform few-layer graphene and promising for various device applications. These include the ambipolar field effect with high carrier mobility and quasi-2D transport with quantum coherence. The fact that these promising electronic properties can still be observed in such films despite their non-uniformity, and thus are not restricted to uniform few-layer graphene, can be important for employing various graphene-based materials produced by diverse methods in practical applications ranging from thin film transistors, high speed/high frequency devices[35,36] to conductive coatings, transparent electrodes[49] or flexible electronics[9].

**Acknowledgements.** YPC acknowledges support from Miller Family Endowment, Birck Director's Fund and Semiconductor Research Corporation (SRC)'s Nanoelectronics Research Initiative (NRI) via Midwest Institute for Nanoelectronics Discovery (MIND). HC acknowledges support from Grodzins endowment. Acknowledgment is also made to the donors of the American Chemical Society Petroleum Research Fund for partial support of this research. QY acknowledges support by NSF Grant 0620906 and CAM Special Funding. A portion of this work was carried out at the National High Magnetic Field Laboratory, which is supported by NSF Cooperative Agreement No. DMR-0084173, by the State of Florida and DOE. We thank Jun-Hyun Park and Eric Palm for experimental assistance. We also thank Xi Chen and Ron Reifenberger for helpful discussions.

**Appendix**

*Synthesis and transfer of graphene layers*

The procedures used to synthesize graphene layers on Ni substrates are detailed in Ref. 7, which also demonstrated a different method from what is used in this work to transfer the graphitic thin films onto other substrates. Briefly, we dissolve carbon (decomposed from precursor gases containing hydrocarbon, such as $CH_4$) into bulk Ni (serving also as a catalyst) foils at ~1000°C, followed by cooling using a carefully chosen rate[7] (~10 °C/s) to room temperature. During the cool-down, the solubility of carbon in Ni decreases and carbon segregates at the surface of the Ni substrate to form graphene layers. Afterwards the Ni substrate with the synthesized graphitic thin film (usually scratched slightly to aid its removal) is soaked in high concentration $HNO_3$ solution. After a few minutes, the film detaches from the substrate and floats on the acid surface, and is ready to be transferred.

*AFM, STM and TEM measurements*

AFM (Nanoscope IIIa, Digital Instruments) topography images (height and phase contrast) are taken with the tapping mode under ambient conditions. STM images are taken under ambient conditions by a Nanotech Electronica's Dulcinea Scanning Probe Microscopy (SPM) system, whose head is housed in a Faraday Cage in a low noise room. The graphitic thin film transferred to $Si/SiO_2$ is connected with thin Cu tapes to provide a conducting path in STM measurements. Graphitic lattices (Fig. 2f) are imaged in the relative flat regions (away from wrinkles) of the sample. The STM images are analyzed using WSxM software version 13.0 and a wavelet-based filter[50] is used to enhance the contrast of atomic images. The atomic images of highly oriented pyrolytic graphite (HOPG) are used to calibrate the X-Y piezo. TEM cross sections (~100nm wide along the imaging direction) were prepared by a focused ion beam (FIB) liftout method, with an initial protective layer deposited from a platinum source under the electron beam. An FEI Titan 80-300 TEM operating at either 300 keV (Figs. 2d, 3c) or 80 keV (Fig. 2e) was used to examine the cross sections. Images were recorded through a Gatan imaging filter, and typically zero loss filtered with a 10 eV window to improve clarity and contrast. The magnified insets in Fig. 2d has been processed using a rolling ball background subtraction filter[51]. The 300keV imaging electron beam and the process used to prepare the cross-section are partially disruptive to the sample (especially for the top graphene layers), making it particularly challenging to study thinner regions (<10 layers) and contributing to the defective features often seen in the top layers of the cross-section TEM images (Figs. 2d, 3c). Using lower voltage (80 keV) imaging beam can help prevent electron beam induced damage at the cost of lower image quality. Interpreting the images of the top graphene layers can also be complicated by the significant out of plane fluctuations (wrinkles) along the line-of-sight of the imaging beam. The sharpness of the wrinkle features (Fig. 3c) is found to be similar in multiple samples and also decrease with prolonged TEM exposure, suggesting that the features are likely intrinsic to the sample rather than artifacts created by TEM.

*Electronic transport measurements and weak localization analysis*

Electronic transport measurements are performed in a variable temperature insert (Scientific Magnetics) and a He-3 system (Janis), using standard low frequency (~10 Hz) lock-in (SRS) detection, with typical excitation current of 100 nA. The WL correction in the low temperature magnetoconductivity (Fig. 5) $\Delta\sigma(B) = \sigma(B) - \sigma(B=0\,\text{T})$ is fitted to the theoretically predicted form[38] for 2D diffusive metals,

$$\Delta\sigma_{WL}(B) = \left(\frac{2e^2}{\pi h}\right)\left(F[\frac{8\pi B}{(h/e)L_\varphi^{-2}}] - F[\frac{8\pi B}{(h/e)L_{tr}^{-2}}]\right)$$

, where $e$ is electron charge, $h$ is the Planck constant, $F[z] = \ln(z) + \Psi(1/2 + 1/z)$ with $\Psi$ being the Digamma function. $L_\varphi$ (phase coherence length) and $L_{tr}$ (transport scattering length) are two fitting parameters.

**References**

(1) Geim, A.K. & Novoselov, K.S. *Nature Mat*. **2007**, *6*, 183

(2) De Heer, W. A; Berger, C.; Wu, X.; First, P. N.; Conrad, E. H.; Li, X.; Li, T.; Sprinkle, M.; Hass, J.; Sadowski, M. L.; Potemski, M.; Martinez, G. *Solid State Commun*. **2007**, *143*, 92-100.

(3) Park, S.; Ruoff, R. S. *Nature Nanotech*. **2009**, *4*, 217-224.

(4) Obraztsov, A.N. *Nature Nanotech*. **2009**, *4*, 212-213.

(5) Oshima, C.; Nagashima, A. *J. Phys. Cond. Mat*. **1997**, *9*, 1-20.

(6) Karu, A E. and Beer, M. *J. Appl. Phys.* **1966,** *37***,** 2179

(7) Yu, Q. K.; Lian. J., Siripongert, S.; Li, H.; Chen, Y. P.; Pei. S. S. *Appl. Phys. Lett*. **2008**, *93*, 113103.

(8) Reina, A.; Jia, X.; Ho, J.; Nezich, D.; Son, H.; Bulovic, V.; Dresselhaus, M. S.; Kong. J. *Nano Lett*. **2009**, *9*, 30-35.

(9) Kim, K.S.; Zhao, Y.; Jang, H.; Lee, S. Y.; Kim, J. M.; Kim, K. S.; Ahn, J. H.; Kim, P.; Choi, J. Y.; Hong, B. H. *Nature* **2009**, *457*, 706-710.

(10) Gomez De Arco, L.; Zhang, Y.; Kumar, A.; Zhou, C. *IEEE Trans. Nanotech*. **2009**, *8*, 135-138.

(11) Chae, S. J.; Gunes, F.; Kim, K. K.; Kim, E. S.; Han, G. H.; Kim, S. M.; Shin, H. J.; Yoon, S. M.; Choi, J. Y.; Park, M. H.; Yang, C. W.; Pribat, D.; Lee, Y. H.; *Adv. Mater.* **2009**, *21*, 1-6.

(12) Obraztsov, A.N.; Obraztsova, E.A.; Tyurnina, A.V.; Zolotukhin, A.A. *Carbon* **2007**, *45*, 2017-2021.

(13) Gilje, S.; Han, S.; Wang, M.; Wang, K. L.; Kaner R.B. *Nano Lett*. **2007**, *7*, 3394-3398.

(14) Gomez-Navarro, C.; Weitz, R. T.; Bittner, A. M.; Scolari, M.; Mews, A.; Burghard, M.; Kern, K.; *Nano Lett*. **2007**, *7*, 3499-3503.

(15) Cote, L.J.; Kim, F.; Huang, J. *J. Am. Chem. Soc.* **2009**, *131*, 1043-1049.

(16) Dikin, D.A.; Stankovich, S.; Zimney, E.J.; Piner, R.D.; Dommett, G.H.B.; Evmenenko, G.; Nguyen, S.T.; Ruoff, R.S. *Nature* **2007**, *448*, 457-460.

(17) Blake, P.; Hill, E. W.; Neto, A. H. C.; Novoselov, K. S.; Jiang, D.; Yang, R.; Booth, T. J.; Geim, A. K. *Appl. Phys. Lett*. **2007**, *91*, 063124

(18) Ni, Z.H.; Wang, H.M.; Kasim, J.; Fan, H.M.; Yu, T.; Wu, Y.H.; Feng, Y.P.; Shen, Z.X. *Nano Lett*. **2007**, *7*, 2758

(19) Novoselov, K. S.; Geim, A. K.; Morozov, S. V.; Jiang, D.; Zhang, Y.; Dubonos, S. V.; Grigorieva, I. V.; Firsov, A. A. *Science* **2004**, *306*, 666-669.

(20) Ferrari, A. C.; Meyer, J. C.; Scardaci, V.; Casiraghi, C.; Lazzeri, M.; Mauri, F.; Piscanec, S.; Jiang, D.; Novoselov, K. S.; Roth, S.; Geim, A. K. *Phys. Rev. Lett*. **2006**, *97*, 187401

(21) Gupta, A.; Chen, G.; Joshi, P.; Tadigadapa, S.; Eklund, P. C. *Nano Lett*. **2006**, *6*, 2667-2273

(22) Graf, D.; Molitor, F.; Ensslin, K.; Stampfer, C.; Jungen, A.; Hierold, C.; Wirtz, L. *Nano Lett*. **2007**, *7*, 238-242.

(23) Meyer, J.C.; Kisielowski, C.; Erni, R.; Rossell, M. D.; Crommie, M. F.; Zettl, A. *Nano Lett*. **2008**, *8*, 3582-3586.

(24) Kollie, T. G. *Phys. Rev. B* **1977**, *16*, 4872-4881.

(25) Pierson H. O. *Handbook of carbon, graphite, diamond and fullerenes--- Properties, Processing and Applications* (Noyes Publications, Park Ridge, **1993**).

(26) Cambaz, Z.G.; Yushin, G.; Osswald, S.; Mochalin, V.; Goyotsi, Y. *Carbon* **2007**, *46*, 841-849.

(27) Freund, L.B.; Suresh, S. *Thin film materials: stress, defect formation and surface evolution* (Cambridge Univ. Press, Cambridge, **2003**)

(28) Kim, D.H.; Ahn, J. H.; Choi, W. M.; Kim, H.S.; Kim, T.H.; Song, J.; Huang, Y. Y.; Liu, Z.; Lu, C.; Rogers, J. A. *Science* **2008** *320*, 507-511.

(29) Nelson, D. R.; Piran, T.; Weinberg, S. *Statistical mechanics of membranes and surfaces* (World Scientific, Singapore, **2004**)

(30) Meyer, J.C.; Geim, A. K.; Katsnelson, M. I.; Novoselov, K. S. ; Booth, T. J. ; Roth, S. *Nature* **2007**, *446*, 60-63.

(31) Fasolino, A.; Los, J.H.; Katsnelson, M.I. *Nature Mater*. **2007**, *6*, 858-861.

(32) Tikhonenko, F. V.; Horsell, D. W.; Gorbachev, R. V.; Savchenko, A. K. *Phys. Rev. Lett*. **2008**, *100*, 056802.

(33) Zhang, Y.; Small, J. P.; Amori, M. E. S.; Kim, P. *Phys. Rev. Lett*. **2005**, *94*, 176803.

(34) Morozov, S.V.; Novoselov, K. S.; Schedin, F.; Jiang, D.; Firsov, A. A.; Geim, A. K. *Phys. Rev. B* **2005**, *72*, 201401.

(35) Lin, Y.M.; Jenkins, K. A.; Valdes-Garcia, A.; Small, J. P.; Farmer, D. B.; Avouris, P. *Nano Lett*. **2009**, *9*, 422–426.

(36) Meric, I.; Han, M. Y.; Young, A. F.; Ozyilmaz, B.; Kim, P.; Shepard, K. L. *Nature Nanotech*. **2008**, *3*, 654-659.

(37) Wang, H.; Nezich ,D.; Kong, J.; Palacios, T. *IEEE Electron Device Lett*. **2008**, *5*, 547-549.

(38) Beenakker, C.W.J.; van Houten, H. *Solid State Physics Vol. 44, Chap. 1*, Enrenreich, H. & Turnbull, D. eds. (Academic Press, San Diego, **1991**).

(39) Koike, Y.; Morita, S.; Nakanomyo, T.; Fukase, T. *Journal of the physical society of Japan* **1985**, *54,* 713-724 .

(40) Morozov, S. V.; Novoselov, K. S.; Katsnelson, M. I.; Schedin, F.; Ponomarenko, L. A.; Jiang, D., Geim, A. K.; *Phys. Rev. Lett*. **2006**, *97*, 016801.

(41) Wu, X.; Li, X.; Song, Z.; Berger, C.; de Heer, W. A. *Phys. Rev. Lett*. **2007**, *98*, 136801.

(42) Gorbachev, R. V.; Tikhonenko, F. V.; Mayorov, A. S.; Horsell, D. W.; Savchenko, A. K. *Phys. Rev. Lett*. **2007**, *98*, 176805.

(43) Ki, D.; Jeong, D.; Choi, J.; Lee. H.; Park, K. *Phys. Rev. B* **2008**, *78*, 125409.

(44) Shen, T.; Wu, Y. Q.; Capano, M. A.; Rokhinson, L. P.; Engel, L. W.; Ye, P. D. *Appl. Phys. Lett*. **2008**, *93*, 122102.

(45) Kelly, M.J. *Low-Dimensional Semiconductors: Materials, Physics, Technology, Devices* (Oxford

Univ. Press, Oxford, **1996**).

(46) Datta, S. *Electronic Transport in Mesoscopic Systems* (Cambridge Univ. Press, Cambridge, **1997**).

(47) Emtsev, K.V. *et al. Nature Mater*. **2009**, 8, 203-207.

(48) Li, X.; Cai, W.; An, J.; Kim, S.; Nah, J.; Yang, D.; Piner, R.; Velamakanni, A.; Jung, I.; Tutuc, E.; Banerjee, S. K.; Colombo, L.; Ruoff, R. S. *Science* **2009**, *324*, 1312-1314.

(49) Wang ,X.; Zhi, L.;Müllen, K. *Nano Lett*. **2008**, 8, 323-327.

(50) Gackenheimer, C., Cayon, L., Relfenberger, R., *Ultramicroscopy*. **2006**, 106, 389

(51) Rasband, W.S., ImageJ, U. S. National Institutes of Health, Bethesda, Maryland, USA, http://rsb.info.nih.gov/ij/, 1997-2009